\documentclass[twocolumn,showpacs,preprintnumbers,amsmath,amssymb,aps]{revtex4}

\usepackage{graphicx}
\usepackage{dcolumn}
\usepackage{bm}

\begin{document}

\preprint{}

\title{Complex Hybrid Inflation and Baryogenesis}

\author{David Delepine}
 \email{delepine@fisica.ugto.mx}
\author{Carlos Mart\'inez}
 \email{crmtz@fisica.ugto.mx}
\author{L. Arturo Ure\~na--L\'opez}%
 \email{lurena@fisica.ugto.mx}
\affiliation{Instituto de F\'isica de la Universidad de Guanajuato,
 C.P. 37150, Le\'on, Guanajuato, M\'exico.}%

\date{\today}

\begin{abstract}
We propose a hybrid inflation model with a complex waterfall field
which contains an interaction term that breaks the $U(1)$ global
symmetry associated to the waterfall field charge.  We show that the
asymmetric evolution of the real and imaginary parts of the complex
field during the phase transition at the end of inflation translates
into a charge asymmetry. The latter strongly depends on the vev of the
waterfall field, which is well constrained by diverse cosmological
observations.
\end{abstract}

\pacs{98.80.Cq 11.30.Er 11.30.Fs 12.60.-i}
\maketitle

We know only 4\% out of the total material content of the universe is
of baryonic nature, the type of matter we seem to understand thanks to
the Standard Model of Particle Physics and nucleosynthesis process of
the early universe. However, even in this well understood case one
fundamental question remains open: why this baryonic component is
almost completely made of matter and not of antimatter? This is
usually quoted as the \textit{baryonic asymmetry} problem. Sakharov in
Ref.\cite{Sakharov:1967dj} showed that any quantum field theory
could generate a baryonic asymmetry if three conditions are satisfied:
no-conservation of the baryonic charge, CP and C violation, and out-of
equilibrium condition for the universe. The electroweak standard model
cannot produce the baryonic asymmetry of the universe as the condition
to be out of equilibrium cannot be fulfilled at the electroweak phase
transition.

In this paper, we shall present a particular model of hybrid
inflation\cite{Linde:1993cn,Copeland:1994vg} in which a complex
scalar field is the responsible for both the symmetry breaking that
puts an end to inflation and for the production of a baryonic
asymmetry in the early universe. The Lagrangian of our model is CP
conserving for most of inflation, but we shall show that the dynamics
of the waterfall field will generate an effective CP asymmetry at the
very end of inflation.

The use of a complex waterfall field in hybrid inflation has been
studied before in\cite{Felder:2000hj}, where the process of symmetry
breaking, which produces cosmic strings instead of domain walls, was
studied in some detail. The idea of a complex scalar field as the
responsible for generating a baryonic asymmetry has been proposed
before in some other models, see for
instance\cite{Rangarajan:2001yu,Bauer:2005vz}. In these models, the
complex field can be identified with other relevant scalar fields,
like the inflaton and the quintessence fields.

 A similar idea to ours was worked out in\cite{Berezhiani:2001xx},
 where an extra complex field, apart from the standard real fields of
 hybrid inflation, is the responsible of baryogenesis through an
 Affleck-Dine mechanism. Nevertheless, our model is a simpler one as
 it does not require the introduction of extra fields.

To begin with, we consider that during the early inflationary universe
there were two scalar fields, the inflaton $\phi$, and another complex
one $a$ wearing a  charge which we call \emph{baryonic charge}. These
fields are both minimally coupled to gravity, and endowed with a
scalar potential of the form
\begin{eqnarray}
V(\phi,a) &=& \frac{1}{4\lambda^2} \left( M^2 - \lambda^2 |a|^2
\right)^2 + \left( \frac{m^2}{2} + \frac{g^2}{2} |a|^2 \right) \phi^2
\nonumber \\
&& + \frac{\delta}{4} a^2 \phi^2 + \textrm{c.c.} \label{eq:1}
\end{eqnarray}
This potential resembles the hybrid potential for inflation, and for
$\delta = 0$, the complex field could be identified with the
waterfall field of standard hybrid
inflation\cite{Felder:2000hj}. There are six free parameters, with
$\delta=|\delta|e^{i\theta_\delta}$ being a complex constant.

Notice that potential~(\ref{eq:1}) allows the total Lagrangian
to be renormalizable, and is invariant under the discrete symmetry $a
\to -a$. Another $a^4$-term could be added to the potential, but
such term would have a negligible effect in our calculations below.

In order to grasp the dynamics of the fields, we need to know the
critical points of the scalar field potential. As in the standard
case, there are two critical points of physical interest. The first
one is a local maximum, and is located at $\phi=|a|=0$, and the
corresponding value of the scalar potential is $V(0,0)
=M^4/(4\lambda^2)$; this is considered a false vacuum because is
unstable. The critical value of the complex phase of $a$, $\theta$, at
this point cannot be resolved, but it will depend on the initial
conditions and on the particular evolutionary path that may have
carried the waterfall field into $|a|=0$. The second critical point,
which corresponds to the true vacuum of the system, is a global
minimum and is located at $\phi=0$ and $\lambda |a|/M=1$. Again, the
complex phase $\theta$ cannot be resolved but this time because the
global minimum of the system is degenerate.

Even if the presence of the complex parameter $\delta$ does not have
any effect in the location of the critical points, it should be
noticed that $\delta$ does break the $U(1)$ symmetry of the Lagrangian
whenever both the inflaton and waterfall fields are different from
zero.

On the other hand, we notice that the phase of $\delta$ can be removed
from the Lagrangian making a phase redefinition of the waterfall field
$a$. This will also change the dynamical evolution of the fields, but
the quantities of physical interest in our model happen to be
\emph{invariant} under rotations of the complex plane. For simplicity
then, we shall assume hereafter that the $\delta$-coupling is real.

Our primary quantity of interest, the \emph{charge density} associated
to the waterfall field $a$ is given by $n_a \equiv \mathrm{Im}
(a^\ast \dot{a})$, and obeys a Boltzmann equation in the form
\begin{equation}
  \dot{n}_a + 3H n_a = - |\delta| |a|^2 \phi^2 \sin \left( 2\theta
  \right) \, . \label{eq:3}
\end{equation}
Eq.~(\ref{eq:3}) clearly illustrates that any source for the charge
density comes from the potential term with the $\delta$ coupling. But
such source needs non-zero field values in order to modify the charge
density $n_a$.

As for inflation, we will assume the conditions for the so-called
\emph{vacuum inflation}
(VI)\cite{Linde:1993cn,Copeland:1994vg}. First, the inflaton field
accomplishes the constraint $\phi \gg M /g$ before the beginning of
inflation. As we will show below, that constraint is modified in our
extended model, but the value $M/g$ still gives the correct order of
magnitude to start an inflationary stage. In this case, the effective
mass of the waterfall field is positive definite, and this makes $|a|
\to 0$.

Second, the constant term in potential~(\ref{eq:1}) is initially
the dominant one. From the Friedmann equation of cosmology, this means
that the Hubble parameter during inflation is almost constant, its
value given by $H_0 \simeq (\sqrt{2\pi /3 \lambda^2}) (M^2/m_{Pl})$,
where $m_{Pl}$ is the Planck mass. Hence, the second condition for VI
can also be written as $m \ll H_0$.

Once inflation starts, the waterfall field is located near the local
minimum $|a|=0$, whereas the inflaton field obeys the
\emph{slow-roll} equation $3H_0 \dot{\phi} \simeq - m^2 \phi$. The
universe expands almost exponentially with time, and the scale factor
of the universe evolves in the form $R=R_{end} e^{-N}$, where we have
defined the number of e-foldings to the end of inflation as $N
\equiv H_0 (t_{end} -t )$. The subscript '$end$' in all quantities
means their values at the end of inflation. In this approximation, the
solution for the inflaton field is
\begin{equation}
\phi (t) = \phi_{end} \exp \left[ (m^2/3H^2_0) N\right] \,
. \label{eq:4}
\end{equation}
The only possibility to put an end to VI is by the destabilization
of the waterfall field. In the case $\delta =0$, that proceeds as in
standard hybrid inflation with a real waterfall
field\cite{Felder:2000hj}. However, our model presents some additional
features we are to study below.

During inflation, we describe the motion of the complex waterfall
field around the local minimum $|a|=0$ as that of small
\emph{classical} oscillations. In this regime, the waterfall field is
not coupled to the inflaton, so its real and imaginary parts can be
considered a separate pair of coupled and damped harmonic oscillators
of the form
\begin{eqnarray}
  \ddot{Q}_\pm &=& -3H_0 \dot{Q}_\pm - m^2_\pm (\phi) Q_\pm \, ,
  \label{eq:5} \\
  m^2_\pm (\phi) &=& \left( g^2 \pm |\delta| \right) \phi^2 - M^2 \,
  , \nonumber
\end{eqnarray}
where the $\pm$ signs are in correspondence, and the fields $Q_\pm$
are the perturbations of the real and imaginary parts of $a$,
respectively.

As seen in Eqs.~(\ref{eq:5}), there are two special points in the
slow-roll down of the inflaton field for which the masses of the
oscillators vanish, $\phi_\pm = M /\sqrt{g^2 \pm |\delta|}$, and then
$\phi_+ < \phi_-$. Next, we can use the slow-roll inflaton's solution,
Eq.~(\ref{eq:4}), and the assumption $m \ll H_0$, to give the explicit
time-dependence of the mass terms,
\begin{equation}
  m^2_\pm (t) \simeq \frac{M^2 m^2}{3H_0} (t_\pm -t) \, , \label{eq:6}
\end{equation}
where $t_\pm$ are the corresponding times of $\phi_\pm$ such that
\begin{equation}
  \frac{m^2}{3H_0} (t_+ -t_-) = \ln \left( \frac{\phi_-}{\phi_+} \right)
  = \frac{1}{2} \ln \left( \frac{1+|\delta|/g^2}{1-|\delta|/g^2}
  \right) \simeq  \frac{|\delta|}{g^2} \, . \label{eq:7}
\end{equation}
Therefore, the exact solution of Eqs.~(\ref{eq:5}) can be given in
terms of Airy functions as follows
\begin{subequations}
\label{eq:8}
\begin{eqnarray}
  Q_+(t) &=& C_+ R^{-3/2} f_+ [u_+(t)] \, , \label{eq:8a} \\
  Q_-(t) &=& C_- R^{-3/2} f_- [u_-(t)] \, , \label{eq:8b}
\end{eqnarray}
\end{subequations}
where
\begin{subequations}
\label{eq:9}
\begin{eqnarray}
  f_\pm (u) &=& \mathrm{AiryAi}(u_\pm) + C^\ast_\pm
  \mathrm{AiryBi}(u_\pm) \, , \label{eq:9a} \\
  u_\pm(t) &=& \frac{1}{4} \left( \frac{3H_0}{M^2 m^2} \right)^{2/3}
  \left[ 9H^2_0 - 4m^2_\pm (t) \right] \, , \label{eq:9b}
\end{eqnarray}
\end{subequations}
being  $C_\pm$ and $C^\ast_\pm$ arbitrary constants.

The normal modes $Q_\pm$ are destabilized at different times as the
inflaton field rolls down to the origin of coordinates. We notice
that $Q_-$ is destabilized first from the false vacuum once $\phi
\sim \phi_-$, and $Q_+$ is destabilized next at $\phi \sim \phi_+$.
It is at this point that we can considered the system as completely
destabilized, as both oscillation modes now grow and roll down to
the true vacuum.

However, all the results in Eqs.~(\ref{eq:8}) and~(\ref{eq:9}) are
valid under the assumption that inflation proceeds up to $t = t_+$,
and then if $t_+ \lesssim t_{end}$. We can estimate the value of
$t_{end}$ if we assume that for $t > t_-$ the mode $Q_-$ follows the
effective minimum at fixed $\phi$ once the latter falls below
$\phi_-$\cite{Copeland:1994vg}. The end of inflation would happen at
the time the slow-roll parameter $\epsilon (\phi_{end} )\simeq 1$;
from this and after some involved algebra, we obtain the strong
constraint
\begin{equation}
  \frac{\phi_-}{\phi_{end}} \simeq 1 + \frac{\sqrt{\pi}}{2} \left(
  \frac{\phi_-}{m_{Pl}} \right) \, . \label{eq:10}
\end{equation}
This result confirms that inflation ends almost instantaneously soon
after the critical point ($\phi_-$ in our case) is reached,
as described in more detail
in\cite{Copeland:1994vg,Garcia-Bellido:1996qt,Garcia-Bellido:1997wm}.

We shall take for granted that condition~(\ref{eq:10}) is
  satisfied so that $\phi_+ \gtrsim \phi_{end}$, and then that we
  fulfill the aforementioned restriction $t_+ \lesssim
  t_{end}$. Eqs.~(\ref{eq:7}) and~(\ref{eq:10}) can be combined
  together to impose an \textit{upper limit} on $|\delta|/g^2$, whose
  maximum value is reached for $\phi_+=\phi_{end}$. We assume
  hereafter that the latter condition applies, and then
\begin{equation}
\frac{|\delta|}{g^2} \simeq \ln \left( \frac{\phi_-}{\phi_+} \right)
\simeq \frac{\sqrt{\pi}}{2}\frac{\phi_-}{m_{Pl}} \, .
\end{equation}

In passing, we would like to mention that the spectrum of
primordial perturbations is not sensitive to the complex nature of the
waterfall field $a$, as the final output is under the control of the
inflaton field only. Actually, the amplitude of primordial
perturbations for the potential~(\ref{eq:1}) reads $\tilde{\delta}_H =
\delta_H \sqrt{1-|\delta|/g^2}$, where $\delta_H$ is the standard
hybrid inflation result\cite{Linde:1993cn}. Thus, under the assumption
$|\delta|/g^2 \ll 1$, our model preserves the main results of standard
hybrid inflation as regards density
perturbations\cite{Linde:1993cn,Copeland:1994vg,Garcia-Bellido:1997wm,Garcia-Bellido:1996ke,deVega:2006hb,Spergel:2006hy}.

On the other hand, Eqs.~(\ref{eq:5}) are also the equations of motion
for the quantum fluctuations of the waterfall field. It has been shown
in\cite{Garcia-Bellido:1996qt}, that these quantum fluctuations are
negligible if they are very massive, i.e. $m_\pm > H_0$. But once
their masses are smaller than $H_0$, which happens around the times at
which the modes become massless, $m_\pm = 0$, the fluctuations acquire
an amplitude of the order $Q_\pm \sim H_0/2\pi$. Such an assumption is
not exact, but it is a reasonable one because the contributions to the
full amplitude made by different wavelenghts are comparable; for
further details see\cite{Garcia-Bellido:1996qt}. Therefore, in order
to fix the arbitrary constants in Eqs.~(\ref{eq:8}), we impose the
conditions\cite{Bastero-Gil:1999fz,Berezhiani:2001xx}
\begin{equation}
Q_\pm (t_\pm) \simeq \pm \frac{H_0}{2\pi} \, , \quad \dot{Q}_\pm (t)
\simeq -\frac{1}{3H_0} \frac{\partial V}{\partial Q_\pm} \,
. \label{eq:11}
\end{equation}
These conditions represent the slow-roll motion of the $Q$-fields
close to their instability points.

Notice that the normal modes $Q_\pm$ are evaluated at the points they
 are effectively massless, and that the slow-roll condition suggests
 $\dot{Q}_\pm (t_\pm) \simeq 0$. The latter also implies that
 $C^\ast_+ = C^\ast_-$\footnote{The condition $C^\ast_+ = C^\ast_-$
 also suggests that the initial conditions imposed on $Q_\pm$ at $t
 \ll t_\pm$ were \emph{symmetrical}. As we shall show, our
 model can generate a baryonic asymmetry even if there is none induced
 by initial conditions.}, and then both normal modes $Q_\pm$ can be
 described by the same function,  $f_- = f_+ = f$. Thus,
 Eqs.~(\ref{eq:8}) now read
\begin{subequations}
\label{eq:12}
  \begin{eqnarray}
    Q_+(t) &=& \pm \frac{H_0}{2\pi} \left( \frac{R_+}{R} \right)^{3/2}
    \frac{f[u_+(t)]}{f[u_+(t_+)]} \, , \label{eq:12a} \\
    Q_-(t) &=& \pm \frac{H_0}{2\pi} \left( \frac{R_-}{R} \right)^{3/2}
    \frac{f[u_-(t)]}{f[u_-(t_-)]} \, . \label{eq:12b}
  \end{eqnarray}
\end{subequations}
in which $R_\pm$ indicates the values of the scale factor at which the
respective mode becomes massless.

Any charge asymmetry in the waterfall field should happen in between
the times $t_\pm$, where both scalar fields $\phi$ and $a$ have enough
amplitude to be a source for the Boltzmann equation~(\ref{eq:3}). That
the produced charge is not null critically depends on the asymmetric
evolution of the oscillation modes $Q_\pm$ which allows to generate a
CP asymmetry during the time interval $\Delta t = t_+-t_-$; for
$\delta =0$ the modes are destabilized at the same time and no charge
would be generated. The final result after Eqs.~(\ref{eq:11})
and~(\ref{eq:12}) is then
\begin{eqnarray}
  \left| n_a \right| & \simeq & \dot{Q}_- (t_+) Q_+ (t_+) =
  (1/6\pi) m^2_-(t_+) Q_- (t_+) \, , \nonumber \\
  & \simeq & \frac{M^2 m^2}{54\pi^2 H_0} \left(
  \frac{9H^2_0}{2m^2} \frac{|\delta|}{g^2} \right) \exp \left(
  -\frac{9H^2_0}{2 m^2} \frac{|\delta|}{g^2} \right) \frac{f
    [u_-(t_+)]}{f[u_-(t_-)]} \, .   \label{eq:14}
\end{eqnarray}
Because the proximity of $u_-$ and $u_+$, the last ratio on the
  r.h.s. of Eq.~(\ref{eq:14}) is of order of unity plus corrections
  of order $|\delta|/g^2$; as the asymmetry is already proportional to
  the latter, we will take it equal to unity.

With the help of Eqs.~(\ref{eq:7}) and~(\ref{eq:10}), the argument in
the exponential term in Eq.~(\ref{eq:14}) can be related to the COBE
normalization
condition\cite{Copeland:1994vg,Garcia-Bellido:1996ke,Garcia-Bellido:1997wm},
so that
\begin{equation}
  \frac{9H^2_0}{2 m^2} \frac{|\delta|}{g^2} = \frac{3
  \pi^{3/2}}{2\lambda^2 g} \frac{(M/m_{Pl})^5}{(m/m_{Pl})^2} \simeq
  2.93 \times 10^{-4} \frac{\lambda}{g^2} \, , \label{eq:15}
\end{equation}
where the very last equality appears from the known CMB constraint on
primordial perturbations, see Eq.~(12)
in Ref.\cite{Garcia-Bellido:1997wm}.

Eq.~(\ref{eq:15}) is a fundamental result, as it relates the
asymmetric evolution of modes $Q_{\pm}$ with the amplitude of
inflationary density perturbations. However, we should stress out that
such a relation is only valid if $|\delta| \neq 0$. Also, we would
like to stress out that our assumption $|\delta|/g^2 \ll 1$ is
guaranteed if $m \ll H_0$, see Eq.~(\ref{eq:15}). In consequence, all
our calculations are valid within the regime expected for vacuum
inflation.

To make an order of magnitude estimation, we will assume that
reheating occurs promptly at the end of inflation so that the
reheating temperature is $T_{reh} = [30 /(4\pi^2 g_\ast
  \lambda^2)]^{1/4}M$. If the total entropy is produced in the
reheating stage, the baryonic asymmetry produced in the model
presented here is estimated to be
\begin{equation}
  \left| \frac{n_a}{s} \right| \simeq 1.77 \frac{g^{3/4}_\ast}{q_\ast}
  \frac{M^2}{m^2_{Pl}} x e^{-x^2} \, , \label{eq:16}
\end{equation}
where variable $x^2$ in the exponential is the term given in
Eq.~(\ref{eq:15}), and $g_\ast$ ($q_\ast$) represents the density
(entropy) degrees of freedom, respectively.

Let us take that $g_\ast = q_\ast \simeq 10^2$, and estimate the
\emph{maximum} charge asymmetry carried by the waterfall field; this is
accomplished for $x=1/\sqrt{2}$. If the expected asymmetry is of order
$\sim 10^{-10}$, then we get the \emph{lower} bound $(M/m_{Pl}) >
10^{-5}$. On the other hand, the combined constraints on the amplitude
and on the spectral index of primordial perturbations give the
\emph{upper} bound $(M/m_{Pl}) < 5 \times 10^{-5}
\lambda/g$\cite{Garcia-Bellido:1997wm}.

It is not easy to satisfy both constraints in the general case. We
should note that $\lambda \ll g$ would not be allowed, and so the
constraints seem to prefer $\lambda \gtrsim g$. Actually, all parameters
can be resolved in the case $g \sim \lambda$, for which we get
$(M/m_{Pl}) \sim 10^{-5}$ and $\lambda \sim g \sim 10^{-4}$. This
scenario may not be the most realistic, as the vev of the waterfall
field is just below the Planck scale, and the inflaton field should be
of the same order before the beginning of inflation. In all other
cases, we can summarize the results by normalizing $\lambda =1$, and
get the following range for the remaining free parameter $0.05
< g < 0.5$; correspondingly we get $10^{-5} < (M/m_{Pl}) < 10^{-2}$.

Standard calculations\cite{Kolb:1990vq} show that cosmic
  strings in our model would have a string tension $\mu$ of the order
  of $G\mu \sim (M/\lambda m_{Pl})^2$. Current constraints coming from
  CMB observations suggest $G\mu <
  10^{-6}$\cite{Bevis:2007gh,Sakellariadou:2007bv}. If taken into
  account, this constraint will shorten the allowed range to $10^{-5}
  < (M/m_{Pl}) < 10^{-4}$; this means our model could be falsified if
  cosmological constraints are improved in the near future.

The charge of the waterfall field can be easily translated in a
baryonic asymmetry once the $a$ field decays into fermions,
\begin{equation}
  \left| \frac{n_B}{s} \right| \simeq \kappa \Delta B \frac{\Gamma_{\Delta
  B}}{\Gamma_a} \left| \frac{n_a}{s} \right| \, , \label{eq:17}
\end{equation}
where $\Gamma_a$ is the $a$-total width, $\Gamma_{\Delta B}$ is
the rate of $a$-decays into fermions for a given $\Delta B$, and
$\kappa$ is the suppression factor due to interactions of the
waterfall fields with the inflaton. For instance, the channels
violating the charge of the waterfall field without producing fermions
are $\Gamma( a \rightarrow 2\phi)$ or $\Gamma(a+a\rightarrow \phi +
\phi)$, which are proportional respectively to $g^4/\lambda^2$ and
$|\delta|^2$. These channels are suppressed for some of the values
allowed by the constraints discussed above.

The mechanism presented throughout this paper to generate a baryonic
asymmetry is generic and illustrates how such an asymmetry can be
produced at the phase transition of hybrid inflation. The model
appears to be the simplest realization, as there is no need of extra
fields and the term added to the standard potential is not
complicated.

An order of magnitude estimation was made assuming prompt
reheating after the end of inflation. However, in a more realistic
scenario the fields would roll down to the minima of
potential~(\ref{eq:1}) and start to oscillate, leading to an extended
reheating process and maybe to other mechanisms of coherent
baryogenesis\cite{Berezhiani:2001xx,Garbrecht:2003mn}. As for our
model, we do not expect the production of substantial charge asymmetry
in this last stage, mainly because of the smallness of the inflaton
field after inflation.

For simplicity too, we called baryonic charge the one associated to
the complex waterfall field. However, in realistic models, it would be
better to choose another charge (for instance the $B-L$-charge in
place of the $B$-charge) for the $a$-field to avoid any suppression of
the baryonic asymmetry at further stages in the evolution of the
universe. In case the $a$-field charge is the baryonic charge, it is
clear that the vev of the $a$-field in the true vacuum will
spontaneously break the baryonic charge. But, it is important to
notice that by making a judicious choice of the value of the baryonic
charge of the $a$ field (for instance $Q_a \gtrsim 2$), it is possible
to avoid the restrictions on baryonic violating interactions coming
from the experimental limits on proton decays.

Nevertheless, how this asymmetry could be transferred to the observed
matter-antimatter asymmetry, and how to keep it until our present time
are still open questions. Any answer crucially depends on the coupling
of the waterfall field to ordinary matter and the reheating
  mechanism after inflation; this is work under research and will be
reported elsewhere\cite{delepine}.

\acknowledgments{We thank Abdel P\'erez-Lorenzana and Mark Hindmarsh
  for useful comments and discussions. C. M. acknowledges a scholarship
  from CONACYT. This work was also partially supported by CONACYT
  grants (42748, 46195 and 47641), DINPO 85, and PROMEP grants
  UGTO-PTC and UGTO-CA-3.}

\bibliography{charged-hybridrefs}

\end{document}